\documentclass[amsmath,amssymb,nofootinbib,superscriptaddress]{revtex4-1}
\usepackage{color}
\usepackage[dvipsnames]{xcolor}
\usepackage{graphicx,epsfig,hyperref,latexsym}
\usepackage{amsmath}
\usepackage{tikz}
\usetikzlibrary{patterns}
\begin{document}
\title{Regular Friedmann Universes and Matter Transformations}

\author{Alexander Kamenshchik}
\email{kamenshik@bo.infn.it}
\affiliation{Dipartimento di Fisica e Astronomia, Universit\`{a} di Bologna,
via Irnerio 46, 40126 Bologna, Italy\\I.N.F.N., Sezione di Bologna, viale B. Pichat 6/2, 40127 Bologna, Italy}
\author{Polina Petriakova}
\email{polina.petriakova@bo.infn.it}
\affiliation{Dipartimento di Fisica e Astronomia, Universit\`{a} di Bologna,
via Irnerio 46, 40126 Bologna, Italy\\I.N.F.N., Sezione di Bologna, viale B. Pichat 6/2, 40127 Bologna, Italy}

\begin{abstract}\vspace{-0.1cm}
We apply a very simple procedure to construct non-singular cosmological models for flat Friedmann universes filled with minimally coupled scalar fields or by tachyon Born--Infeld-type fields. Remarkably, for the minimally coupled scalar field and the tachyon field, the regularity of the cosmological evolution, or in other words, the existence of bounce, implies the necessity of the transition between scalar fields with standard kinetic terms to those with phantom ones. %Please ensure meaning has been retained. 
In both cases, the potentials %of the minimally coupled scalar field and that of the tachyon 
in the vicinity of the point of the transition have a non-analyticity of the cusp form that is characterized by the same exponent and is equal to $\tfrac23$. If, in the tachyon model's evolution, the pressure changes its sign, then another transformation of the Born--Infeld-type field occurs: the tachyon transforms into a pseudotachyon, and vice versa. We also undertake an analysis of the stability of the cosmological evolution in our models; we rely on the study of the speed of sound squared.
\end{abstract}

\maketitle

\section{Introduction}
{For many years, the cosmological singularity has been one of the most attractive problems in general relativity. Starting from Robertson's seminal work~\cite{Robertson}, the initial singularity issue of Friedmann-type cosmologies was under scrutiny. Launching the singularity-to-maximal-radius-and-back cyclic evolution was already considered there. It seemed that touching the singularity did not bother him too much. Thereafter, generalization to the case of not only homogeneous and isotropic spacetimes was explored \cite{Lif-Khal,Pen,Hawk,Pen-Hawk}, resulting in the proof of some general theorems and the discovery of the oscillatory (BKL) approach to the cosmological initial singularity~\cite{BKL}, also known as the Mixmaster Universe~\cite{Misner}. The investigation of arising (rather soft) future singularities at the finite scale factor was done further~\cite{Barrow} and still maintains interest~\cite{Barrow1,Barrow2,Barrow3,we-tach,Shtan,Kam}. Regarding such soft future singularities, the condition of their crossing becomes important; see, e.g.,~\cite{Laz,we-tach1}. The idea of the possible crossing of the so-called Big Bang{--}Big Crunch singularity appears rather counterintuitive in contrast to the crossing of the soft singularities. Thus, the desire to find models free of such singularities is understandably strong, and this direction is prevalent. However, the idea of the possible transition from a Big Crunch{--}Big Bang transition was also studied in some cosmological models. Let us point out here the string or pre-Big Bang scenario~\cite{string,string1,string2}, wherein the accelerated expansion of the universe is driven by the kinetic energy of the dilaton field. From the cosmological singularity's point of view, its presence is essential, since by making the transition from the string frame---where the dilaton is non-trivially coupled to gravity---to the Einstein frame, the observable evolution of the universe can be drastically changed: what looks like an expansion in one frame can look like a contraction in another one. An alternative treatment is to reformulate the theory, relying on the role of scalar fields, to define the finite variables as the scale factor shrinks to zero. That suggests a natural way to match the solutions before and after the singularity crossing and was inspired by superstring theories~\cite{Khouri,Khouri1,Khouri2}. The importance of such features is significant in other approaches as well~\cite{we-cross,we-cross2,we-cross3,Bars,Bars1, Wett,Wett1}.}

While the activity of the description of crossing a singularity %Please ensure meaning has been retained. 
in cosmology becomes intensive, attempts to find cosmological non-singular models still conserve their charm; see, e.g.,~\cite{modern,modern1,modern2}. Even more active is the search for regular black hole solutions. The study of non-singular black holes started a long time ago~\cite{Bardeen}, and for recent reviews, see~\cite{Spallucci,Sebastiani}. One can write down a singularity-free metric ansatz from the Schwarzschild black hole by a simple substitution of the radial coordinate $r$ as $r \rightarrow \sqrt{r^2+b^2}$, as was proposed by Simpson and Visser~\cite{Simpson-Visser}. That results in the following spacetime:
\begin{equation}
ds^2 = \left(1-\frac{2m}{\sqrt{r^2+b^2}}\right)dt^2-\left(1-\frac{2m}{\sqrt{r^2+b^2}}\right)^{-1}dr^2-\bigl(r^2+b^2\bigr)d\Omega_2^2, %(d\theta^2+\sin^2\theta d\varphi^2),
\label{S-V} \end{equation} where $b$ is a parameter, and the singularity at $r=0$ is replaced by a regular minimum of the area function at $r=0$: %Please ensure meaning has been retained. 
a sphere of radius $b$. If $b > 2m$, this spacetime represents a wormhole with a throat at $r=0$; if $b <2m$, one has a black hole with two horizons at $r=\pm\sqrt{4m^2-b^2}$, and the $b=2m$ case corresponds to an extremal black hole with a single horizon at $r=0$. At the hypersurface $r=0$ in the black hole case, the coordinates change their temporal and spatial assignments, which corresponds to a bounce in one of the two scalar factors of the Kantowski--Sachs universe: the so-called black bounce. Afterward, a Vaidya spacetime \cite{Vaidya}, charged black-bounce spacetimes \cite{charged}, and Kerr black holes \cite{Kerr} were ``regularized'' in this Simpson--Visser spirit. This one-parameter extension \eqref{S-V} is sustained by a phantom scalar field and a magnetic field within nonlinear electrodynamics, as was established in~\cite{Bron}.

Generally, in the majority of works devoted to the construction of regular black holes, one can use the method that many years ago was called the ``G-method'' by Synge \cite{Synge}; see also a recent e-print \cite{Ellis}. Using this method, one chooses a metric, substitutes it into the left-hand side of the Einstein equations, and then sees what happens on the right-hand side. The G-method is opposed to the ``T-method'', for which one chooses the form of the matter in the right-hand side of the Einstein equations and then tries to find the metric that satisfies this system of equations by substituting it into the %Please ensure meaning has been retained. 
left-hand side. The advantage of the G-method consists of the fact that it always works (in contrast with the T-method). Unfortunately, the right-hand side of the Einstein equations that arises as the result of the application of the G-method does not always have some reasonable physical sense that %Please ensure meaning has been retained. 
can be identified with some known fields or other types of matter. The remarkable example of a regular black hole sustained by a minimally coupled phantom scalar field with an explicitly known potential was found in~\cite{Bron-Fabris}. Some properties of this solution were studied in further detail~\cite{we-reg-cosm}, and it was also used \cite{Kam-Pet} in an attempt to construct a regular rotating black hole.

Recently, Bronnikov explored~\cite{Bronnikov} the regularized version of the Fisher solution~\cite{Fisher}, which has been rediscovered many times in different contexts \cite{Fisher1,Fisher2,Fisher3,Fisher4,Fisher5,Fisher6,Fisher7,we}. One can observe an interesting transition from the standard scalar field to the phantom one there. Herein, we can note that the Friedmann cosmological models have a simpler structure than the Schwarzschild-like black holes. That gives some hope that, using an analogy with the Simpson{--}Visser prescription \cite{Simpson-Visser} in cosmology, one can obtain rather simple cosmological solutions with the matter content, which can be analyzed, at least qualitatively. Indeed, this was done in paper \cite{we-reg-cosm}. It was shown that when considering a non-singular metric of the flat Friedmann universe filled with a minimally coupled scalar field, one can find two interesting qualitative features of the model: First, at some moment, the standard scalar field becomes a phantom one, i.e., the kinetic term changes its sign. Second, even if we cannot find an explicit expression for the scalar field potential, we can state that it should be non-analytical, or, more precisely, it should have a cusp. Remarkably, a similar phenomenon was observed in the study of the opportunity of the so-called phantom divide line crossing~\cite{we4,we5}, and the forms of the cusps of the potentials %found in \cite{we-reg-cosm} and those 
coincide. 

As is well known, cosmological models with minimally coupled scalar fields are not the only kind of scalar field models. Some time ago, in the context of string-inspired cosmological models, the so-called tachyon fields were studied~\cite{tach,Padman, Feinstein,Frol-Kof-Star,Gib}. {These tachyon fields arising in string theory~\cite{tach} are not connected with the tachyon particles flying with superluminal velocities \cite{tach-part}. Nevertheless, we shall use the term ``tachyons'', which has already become traditional in the cosmological literature.} These tachyons indeed represent scalar fields with non-trivial kinetic terms of the type that was first studied by Born and Infeld~\cite{Born-Infeld}. {It is interesting to note that the birth of Born--Infeld non-linear electrodynamics was at least in part motivated by the desire to eliminate the singularity of the electric field of a point-like electric charge. Remarkably, sharing with the linear Maxwell electrodynamics the electric--magnetic duality and the physical propagation of waves, the Born--Infeld theory manages to tame the divergence of the Coulomb self-energy~\cite{Ketov}. Indeed, the expression for the electric field of the point-like charge $Q$ has the form
\begin{equation}
\vec{E} = \frac{Q}{\sqrt{r^4+Q^2}}\vec{e}_{r}.
\label{BI} \end{equation}  
Thus, one has regularization, which in a way reminds one to put ``by hands'' into the Simpson--Visser-like metrics for black holes and cosmological models. However, here in~\eqref{BI}, the charge $Q$ plays the role of both the source of the electric filed and of the regularizing quantity. The effective density of the electric point-like charge acquires a finite radius, which is %Please ensure meaning has been retained. 
connected with the dimensional parameter $b$ in the definition of the Born--Infeld action~\cite{Born-Infeld}. Later, it was discovered that this action appears as an effective action in supersymmetric theories \cite{Ketov1,Ketov2} as well as in string theory \cite{Seiberg}. %, where the Born-Infeld parameter $b$ is related to the fundamental physical parameters. 
The attempts to construct a Born--Infeld-type extension of gravity, despite not being unique and well-motivated, are under investigation; see the recent review~\cite{Born-Infeld-rep}.}

The interest in cosmological models with tachyons was mainly connected with their possible role as a source of dark energy. However, further studies have shown that the presence of non-trivial kinetic terms in these models can imply the appearance of some very unusual properties. For example, a tachyon cosmological model with a particular potential depending on trigonometrical functions was studied, and two interesting phenomena were discovered: the self-transformation of the tachyon field into a pseudo-tachyon field and the appearance of a particular type of soft future cosmological singularity, which was called ``Big Brake'' in~\cite{we-tach}. 
%The model suggested in \cite{we-tach} was further studied in papers \cite{we-tach11,we-tach12,we-tach13,we-tach1} where some additional particular transformations of Born-Infeld-type fields were discussed. The quantum properties of this family of models were discussed in papers~\cite{quant-tach,quant-tach1,quant-tach2}. 
Thus, taking into account the richness of the cosmological models based on the presence of Born--Infeld-type fields, it is interesting to study regular flat Friedmann cosmological universes filled with such fields and to see what kind of effects one can observe there. This is the main goal of the present paper.
Its structure is the following: in the second section, we present known results for a regular flat Friedmann universe filled with a minimally coupled scalar field \cite{we-reg-cosm}; in the third section, we consider a flat Friedmann universe filled with a tachyon field. The last section contains a discussion of the obtained results.

%%%%%%%%%%%%%%%%%%%%%%%%%%%%%%%%%%%%%%%%%%
\section{Regular Friedmann Universes and Scalar Fields}
The well-known exact solution for a flat Friedmann universe with a massless scalar field $\phi$ is
\begin{equation}
ds^2=dt^2-t^{2/3}\bigl(dx_1^2+dx_2^2+dx_3^2\bigr), 
\label{Fried} %\end{equation} \begin{equation}
\qquad \quad \dot{\phi} = \sqrt{\frac23}\frac{1}{t}.
%\label{KG} 
\end{equation}
Hereafter, dots %Please ensure meaning has been retained. 
refer to time derivatives. Following the Simpson--Visser recipe~\cite{Simpson-Visser}, one can write down the regularized one from~\eqref{Fried} as:
\begin{equation}
ds^2 = dt^2 -\bigl(t^2+b^2\bigr)^{1/3}\bigl(dx_1^2+dx_2^2+dx_3^2\bigr).
\label{Fried1} \end{equation}
A straightforward calculation provides us with the Ricci tensor components:
\begin{equation}
R^0_0 = \frac{2t^2-3b^2}{3\bigl(t^2+b^2\bigr)^4},
%\label{Ricci} \end{equation} \begin{equation}
\qquad R_1^1=R_2^2=R_3^3=-\frac{b^2}{3\bigl(t^2+b^2\bigr)^2},
\label{Ricci1} \end{equation}
and the Ricci scalar:
\begin{equation}
R = \frac{2t^2-6b^2}{3\bigl(t^2+b^2\bigr)^2}.
\label{Ricci2} \end{equation}
Then, the Einstein equations immediately afford the expressions for the energy density and the isotropic pressure of matter as
\begin{equation}
\rho=\frac{t^2}{3\bigl(t^2+b^2\bigr)^2}, \qquad 
%\label{energy} \end{equation} \begin{equation}
p = \frac{t^2-2b^2}{3\bigl(t^2+b^2\bigr)^2}.
\label{pressure} \end{equation}
Considering spacetime that is filled with a spatially homogeneous scalar field with some potential $V(\phi)$, namely, 
\begin{equation}
\rho = \frac12\dot{\phi}^2+V(\phi), \qquad 
%\label{energy1} \end{equation} \begin{equation}
p = \frac12\dot{\phi}^2-V(\phi),
\label{pressure1} \end{equation}
one can compare these expressions and gain 
\begin{align}\label{scalar} 
\dot{\phi} &= \pm\sqrt{\frac23}\frac{\sqrt{t^2-b^2}}{\, \, t^2+b^2}, \\
%\end{align} and \begin{equation}
V &= \frac{b^2}{3\bigl(t^2+b^2\bigr)^2}.
\label{poten} \end{align}
Equation~\eqref{scalar} can be integrated, providing the field $\phi$ as a function of time $t$. However, we are not able to invert the result and find $t$ as an explicit function of $\phi$, and thus we cannot use Equation~\eqref{poten} to find the explicit form of the potential in terms of the scalar field. Nonetheless, the Formulas \eqref{scalar} and \eqref{poten} provide us with rather interesting information. One can see that the expression \eqref{scalar} makes sense only if $|t| \geq b$. {What would happen at $|t | < b$?} In this situation, the kinetic energy of $\phi$ changes sign, and the standard scalar field transition to a phantom one appears. Therefore, one can observe an analogous effect to that explored in \cite{Bronnikov}. 
%As far as the form of the potential is concerned, we can study its 
The behavior in the vicinity of $t = b$ can be defined through $t = b+\tau$, $ \tau << b $,
%\label{time} \end{equation}
%where $\tau$ is small Then, from Equation \eqref{scalar} with the choice of an overall plus sign (which is not essential), we obtain
resulting in
\begin{equation}
\frac{d\phi}{d\tau} = \frac{\sqrt{\tau}}{\sqrt{3b^3}}, \quad \rightarrow \quad \phi(\tau) = \phi_0+\frac{2\tau^{3/2}}{3\sqrt{3b^3}},
%\label{scalar1} \end{equation} and  \begin{equation} \phi(\tau) = \phi_0+\frac{2\tau^{3/2}}{3\sqrt{3b^3}}, \label{scalar2}
\end{equation} where $\phi_0$ is an integration constant. 
Accordingly,
\begin{equation}
\tau = 3b\left(\frac{\phi-\phi_0}{2}\right)^{2/3}.
\label{time1}
\end{equation} 
Replacing in the expression the potential with %\eqref{time} and 
\eqref{time1}, one can find the behavior near the vicinity of the critical point, and by keeping the leading terms, we have
\begin{equation}
V(\phi) = \frac{1}{12b^2}\left(1-6\left(\frac{\phi-\phi_0}{2}\right)^{2/3}\right).
\label{poten2} \end{equation}
The presence of cusp type's non-analyticity in the expression above is responsible for the transition from the standard scalar field to its phantom counterpart and vice versa.    

One can also consider a slightly more general model:
\begin{equation}
ds^2=dt^2-t^{2\alpha}\bigl(dx_1^2+dx_2^2+dx_3^2\bigr),
\label{Fried2} \end{equation}
in which the dynamics evolve a perfect fluid with the equation-of-state parameter as follows:
\begin{equation}
w = \frac{2-3\alpha}{3\alpha}.
\label{eq-state} \end{equation}
This is the well-known particular solution for the flat Friedmann model with a minimally coupled scalar field and exponential potential. To eliminate the cosmological singularity, one can modify metric~\eqref{Fried2} in a Simpson--Visser spirit as
\begin{equation} 
ds^2=dt^2-\bigl(t^2+b^2\bigr)^{\alpha} \bigl(dx_1^2+dx_2^2+dx_3^2\bigr);
\label{Fried3} \end{equation}
the corresponding Ricci tensor components are
\begin{equation}
R_0^0 = -\frac{3\alpha\bigl((\alpha-1)t^2 + b^2\bigr)}{\bigl(t^2+b^2\bigr)^2}, \qquad 
\label{Ricci3}
%\end{equation} \begin{equation}
R_1^1=R_2^2=R_3^3=-\frac{\alpha\bigl((3\alpha-1)t^2+b^2\bigr)}{\bigl(t^2+b^2\bigr)^2}, %\label{Ricci4}
\end{equation}
and the Ricci scalar is 
\begin{equation}
R=-\frac{6\alpha\bigl((2\alpha-1)t^2+b^2\bigr)}{\bigl(t^2+b^2\bigr)^2}.
\label{Ricci5} \end{equation}
Now the expressions for energy density and pressure read 
\begin{equation}
\rho=\frac{3\alpha^2t^2}{\bigl(t^2+b^2\bigr)^2} , \qquad 
\label{energy2}
%\end{equation} \begin{equation}
p=-\frac{\alpha\bigl((3\alpha-2)t^2+ 2b^2\bigr)}{\bigl(t^2+b^2\bigr)^2},
%\label{pressure2}
\end{equation}
and the corresponding expressions for the potential and the time derivative of the scalar field realizing the evolution \eqref{Fried3} are
\begin{equation}
V(\phi) = \frac{\alpha\bigl((3\alpha-1)t^2+b^2\bigr)}{\bigl(t^2+b^2\bigr)^2}, \qquad 
\label{poten3} 
\dot{\phi}^2 =\frac{2\alpha\bigl(t^2-b^2\bigr)}{\bigl(t^2+b^2\bigr)^2}. \end{equation} 
In the absence of the regularizing parameter $b=0$, we can get from Equation \eqref{poten3} the known expression for the exponential potential:
\begin{equation}
V(\phi) = \alpha\bigl(3\alpha-1\bigr)\exp\left(-\sqrt{\frac{2}{\alpha}}\bigl(\phi-\phi_0\bigr)\right).
\label{poten4} \end{equation}
Nevertheless, if $b > 0$, one can see that, just as in the previous case, the transition from the standard scalar field to the phantom one (or vice versa) takes place. Now, we can again consider the vicinity of the instant $t=b$. Proceeding in a similar way, we obtain the following expression for the behavior of the potential in the vicinity of the cusp:
\begin{equation}
V(\phi) = \frac{\alpha}{4b^2}\left(3\alpha-\frac{2\cdot3^{2/3}}{\alpha^{1/3}}\left(\frac{\phi-\phi_0}{2}\right)^{2/3}\right).
\label{poten5}
\end{equation}
This expression has the same non-analyticity ($\sim$$(\phi-\phi_0)^{2/3}$) as that seen in the expression \eqref{poten2}, and when $\alpha = \tfrac13$, these expressions coincide.

\section{Regular Friedmann Universes and Tachyons}

Let us now again consider a regular flat Friedmann universe with the metric \eqref{Fried3}. The expressions for the components of the Ricci tensor, Ricci scalar, energy, and pressure are given by Equations \eqref{Ricci3}--\eqref{energy2}. However, now the universe is filled by the tachyon (Born--Infeld-type) field with the Lagrangian \cite{tach}:
\begin{equation}
L = -V(T)\sqrt{1-T_{,\mu} T^{,\mu}\, }
\label{tach}
\end{equation} 
where $T$ is the tachyon field, and the function $V(T)$ will be called the ``potential'' of the tachyon field. In the framework of our Friedmann model, we shall consider a spatially homogeneous tachyon field $T = T(t)$, and the Lagrangian \eqref{tach} will take the simple form
\begin{equation}
L = -V(T)\sqrt{1-\dot{T}^2}.
\label{tach1}
\end{equation} 
The energy density and the pressure for this field are 
\begin{equation}
\rho = \frac{V(T)}{\sqrt{1-\dot{T}^2}}, \qquad 
\label{energy-tach}
p = -V(T)\sqrt{1-\dot{T}^2}.
%\label{pressure-tach}
\end{equation} 
The analogue of the Klein--Gordon equation now looks as follows:
\begin{equation}
\frac{\ddot{T}}{1-\dot{T}^2}+\frac{3\alpha t}{\bigl(t^2+b^2\bigr)}\dot{T}+\frac{V_{,T}}{V}=0.
\label{KG-tach} \end{equation}
Comparing the expressions \eqref{energy-tach} %and \eqref{pressure-tach} 
for the tachyon field with the corresponding components of the energy--momentum tensor coming from Friedmann's equations \eqref{energy2}, we obtain
\begin{equation}
\dot{T}^2= \frac{\rho+p}{\rho} = \frac{2\bigl(t^2-b^2\bigr)}{3\alpha t^2},
\label{kin-tach} \end{equation} \begin{equation}
V(T) = \sqrt{-\rho p} = \frac{\sqrt{3\alpha^3t^2\Bigl((3\alpha-2)t^2+2b^2\Bigr)}}{\bigl(t^2+b^2\bigr)^2}.
\label{pot-tach} \end{equation}
One can solve Equation \eqref{kin-tach}, to find a solution
\begin{equation}
T(t) = T_0\pm \sqrt{\frac{2\bigl(t^2-b^2\bigr)}{3\alpha}}\left(1-\frac{b}{\sqrt{t^2-b^2}}\, {\rm arctan}\frac{\sqrt{t^2-b^2}}{b}\right).
\label{tach-sol} \end{equation}

Let us note here that this solution automatically satisfies %the Klein-Gordon equation 
Equation~\eqref{KG-tach} due to the Bianchi identities. This feature is typical for the reconstruction techniques for the potentials of both the minimally coupled and the tachyon fields; see, e.g., ref. %MDPI: Newly added  information. Please confirm.
 \cite{we-tach} and the references~therein. We cannot invert Equation \eqref{tach-sol} and find the cosmic time parameter $t$ as a function of the tachyon field $T$. Thus, as a result, we cannot find an explicit form of the tachyon potential \eqref{pot-tach} as a function of $T$. Let us compare this situation with that of the singular cosmology for which the regularizing parameter $b=0$. In this case,
the universe expands (or contracts) following a simple power law, and Equations \eqref{kin-tach} and \eqref{pot-tach} become simpler:
\begin{equation}
\dot{T}^2= \frac{\rho+p}{\rho} = \frac{2}{3\alpha} \, , \qquad 
\label{kin-tach1} %\end{equation} \begin{equation}
V(T) = \sqrt{-\rho p} = \frac{\sqrt{3\alpha^3\bigl(3\alpha-2\bigr)}}{t^2}.
%\label{pot-tach1} 
\end{equation}
Integrating Equation \eqref{kin-tach1}, we get
\begin{equation}
T(t) = T_0 \pm \sqrt{\frac{2}{3\alpha}}t,  
\label{tach-sol1} \end{equation}
and inverting Equation \eqref{tach-sol1}, one obtains
\begin{equation}
t = \pm \sqrt{\frac{3\alpha}{2}}\bigl(T-T_0\bigr).
\label{tach-sol2} \end{equation}
Substituting expression \eqref{tach-sol2} into Equation \eqref{kin-tach1}, we find the explicit form of the tachyon~potential:
\begin{equation}
V(T) =\frac{\sqrt{\dfrac{4\alpha(3\alpha-2)}{3}}}{\bigl(T-T_0\bigr)^2}.
\label{pot-tach2} \end{equation}

A tachyon model with potential \eqref{pot-tach2} was considered in papers \cite{Padman,Feinstein}. Such a model has a particular exact solution that describes a universe expanding according to the power law with a negative effective pressure. In our terms, it corresponds to the parameter $\alpha$ such that $\alpha > \frac23$. To have a flat Friedmann universe expanding according to the power law but with positive pressure, i.e., with the parameter $\alpha < \frac23$, one can introduce another type of the Born--Infeld-type field, which is called a ``pseudotachyon'' and is described by the following Lagrangian~\cite{we-tach}: %, which for the spatially homogenous field looks like
\begin{equation}
L = V(T)\sqrt{\dot{T}^2-1}.
\label{Lagrange-pseudo}
\end{equation}   
Furthermore, it was shown that it is possible to construct a  potential of the tachyon field with the Lagrangian \eqref{tach1} such that the dynamics drive the universe to the point where the transformation of the tachyon field into a pseudotachyon field is unavoidable and arises in a natural way.

Let us come back to a flat Friedmann universe with metric \eqref{Fried3} and non-singular evolution, i.e., with $b > 0$. We shall first consider the model with $\alpha > \frac23$. In this case, the pressure is always negative, and the expression for the potential \eqref{pot-tach} is well defined. 
Using the obtained expression \eqref{kin-tach}, one can find that 
\begin{equation}
\sqrt{1-\dot{T}^2} = \sqrt{1-\frac{3}{3\alpha}+\frac{b^2}{3\alpha t^2}}
\label{kin-tach2} \end{equation}
is also well defined at $\alpha > \frac23$. However, we see that at $|t| < b$, the right-hand side of Equation \eqref{kin-tach} becomes negative. That means that at the moment in time when %Please ensure meaning has been retained. 
$t = \pm b$, we encounter the transformation of the tachyon field into the phantom tachyon field with the~Lagrangian:
\begin{equation}
L = -V(T) \sqrt{1+\dot{T}^2}.
\label{Lagrange-phant}
\end{equation}
Thus, the universe at $|t| > b$ is driven by the tachyon field, while at $|t| < b$, it is driven by the phantom tachyon field. Note that the transformation between these two types of Born--Infeld-type fields also occurs if $\alpha < \frac23$. 

It is interesting to look at the form of the potential in the vicinity of the point of this transition using the same method that was used in the preceding section for the analysis of the models with minimally coupled scalar fields. Straightforward calculations show that in the vicinity of the phantom--non-phantom transition point, the potential has the following behavior:
\begin{equation}
V = \frac{3\alpha^2}{b^2}-\frac{3\alpha\bigl(3\alpha+1\bigr)}{b^3}\left(\frac{\alpha b}{2}\right)^{1/2}\bigl(T-T_0\bigr)^{2/3}.
\label{pot-tach3} \end{equation}
Note that we again have the same exponent $\frac23$ as in Equation \eqref{poten2} for the potential of the scalar field. 

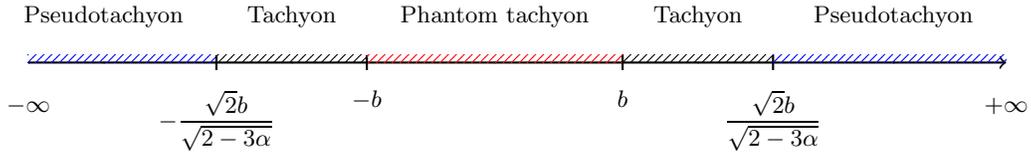
\begin{figure}[th!] \centering
\begin{tikzpicture}[mydrawstyle/.style={draw=black, thick}, x=1mm, y=1mm, z=1mm]
\draw[mydrawstyle, ->](-80,30)--(50,30) node[below=10]{$ +\infty$};
\draw[mydrawstyle](-80,30)--(-80,30) node[below=10]{$-\infty$};
\draw[mydrawstyle](-70,30)--(-70,30) node[above=10]{$\text{Pseudotachyon}$};
\draw[mydrawstyle](-55,29)--(-55,31) node[below=10]{$-\dfrac{\sqrt{2}b}{\sqrt{2-3\alpha}}$};
\draw[mydrawstyle](-45,30)--(-45,30) node[above=10]{$\text{Tachyon}$};
\draw[mydrawstyle](-35,29)--(-35,31) node[below=10]{$-b$};
\draw[mydrawstyle](-18,30)--(-18,30) node[above=10]{$\text{Phantom tachyon}$};
\draw[mydrawstyle](-1,29)--(-1,31) node[below=10]{$b$};
\draw[mydrawstyle](9,30)--(9,30) node[above=10]{$\text{Tachyon}$};
\draw[mydrawstyle](35,30)--(35,30) node[above=10]{$\text{Pseudotachyon}$};
\draw[mydrawstyle](19,29)--(19,31) node[below=10]{$\dfrac{\sqrt{2}b}{\sqrt{2-3\alpha}}$};
\fill[pattern=north east lines,pattern color=blue] (-80,30) rectangle (-55,31);
\fill[pattern=north east lines] (-55,30) rectangle (-35,31);
\fill[pattern=north east lines, pattern color=red] (-35,30) rectangle (-1,31);
\fill[pattern=north east lines] (-1,30) rectangle (19,31);
\fill[pattern=north east lines, pattern color=blue] (19,30) rectangle (50,31);  
\end{tikzpicture}
\caption{Possible transitions between different regimes in the tachyon model \eqref{tach} with $\alpha < \frac23$.} \label{fig:transitions}
\end{figure}

In the case when $\alpha < \frac23$, we have a couple of additional particular time moments
\begin{equation}
t = \pm \sqrt{\frac{2}{2-3\alpha}}\, b
\label{partic} \end{equation}
during which both the expression under the square root in the formula for the potential \eqref{pot-tach} and the expression under the square root for the kinetic structure 
\eqref{kin-tach2} change their signs. This situation is exactly as described in~\cite{we-tach}, and it corresponds to the transition from the tachyon field to the pseudotachyon one. Below, Figure~\ref{fig:transitions} graphically represents the transitions between different regimes in the model with $\alpha < \frac23$. It is easy to see that for $\alpha \geq \frac23$, the transition from tachyon to pseudotachyon is absent.

%\begin{center}
%\end{center}

{It is well known that cosmological solutions avoiding singularities, i.e., solutions with bounces, suffer from instability. While detailed analysis of cosmological perturbations represents a rather cumbersome task that lies beyond the scope of the present paper, we can undertake the study of the speed of sound squared for a cosmological model with the metric \eqref{Fried3}. This analysis will be relevant for both the scalar model of the preceding section and the tachyon model. %The sound speed squared is given by the formula \begin{equation} c_s^2 = \frac{dp}{d\rho}. \label{sound} \end{equation}
We have the expressions for the time dependencies of the pressure and energy densities with respect to time; see Equations~\eqref{pressure} and ~\eqref{energy2}. 
Using these expressions, one can find 
\begin{equation}
c_s^2 = \frac{dp^\prime_{t^2}}{d\rho^\prime_{t^2}} %= \frac{\frac{dp}{d(t^2)}}{\frac{d\rho}{d(t^2)}} 
= \frac{(2-3\alpha)t^2 -3(2-\alpha)b^2}{3\alpha(t^2 -b^2)}.
\label{sound1} \end{equation}
Now we are able to study the time behavior of the speed of sound squared for models with different values of the parameter $\alpha$, characterizing our cosmological evolution.}

{Let us start with the case $\alpha > 2$, which, in the model with the non-regularized metric, i.e., at $b=0$, corresponds to a Friedmann universe filled with a perfect fluid with negative pressure and an equation-of-state parameter $w \leq -\frac23$. First of all, we note that at all values of parameter $\alpha$, the denominator of the expression \eqref{sound1} is positive at $t^2 > b^2$ and negative at $t^2 < b^2$. For $\alpha > 2$, by catching the sign of the numerator, one can easily see that 
\begin{equation}
c_s^2 < 0 \,\,\,  {\rm if} \,\,\, t^2 > b^2 \,\,\, \text{or} \,\,\, t^2 < \frac{3(2-\alpha)}{2 - 3\alpha}\,  b^2, \quad \text{while} \quad c_s^2 > 0 \,\,\,  {\rm if} \,\,\, \frac{3(\alpha - 2)}{3\alpha-2}b^2 < t^2 < b^2.  \label{sound3} \end{equation}
Then, if the time belongs to interval~\eqref{sound3}, where the speed of sound squared is positive, we may ask ourselves when $c_s^2$ is subluminal and when it is superluminal. A simple analysis shows that 
\begin{equation} c_s^2 < 1 \quad  {\rm if} \quad \frac{3(\alpha - 2)}{3\alpha-2}b^2 < t^2 < \frac{3(\alpha - 1)}{3\alpha-1}b^2; \quad \text{and if} \quad \frac{3(\alpha - 1)}{3\alpha-1}b^2 < t^2 < b^2, \end{equation}
one has a superluminal velocity for the propagation of the perturbations.}

{One can consider the case $1 < \alpha \leq 2$, which matches a perfect fluid with a negative pressure with the equation-of-state parameter $-\frac23 \leq w < -\frac13$. A similar analysis shows that $ c_s^2 < 0$ at $t^2 > b^2$. The speed of sound squared is positive, subluminal, and superluminal, correspondingly, at
\begin{equation}
t^2 < \frac{3(\alpha-1)}{3\alpha-1}b^2 \quad \text{and} \quad  \frac{3(\alpha-1)}{3\alpha-1}b^2 < t^2 < b^2. \end{equation}}

{The next case is $\frac23 \leq \alpha \leq 1$, which agrees with a perfect fluid with negative pressure and equation-of-state parameter $-\frac13 \leq w \leq 0$, and $ c_s^2 < 0$ at $t^2 > b^2$ and positive, but it is superluminal at $t^2 < b^2$.}
 
The case $\frac13 \leq \alpha < \frac23$ corresponds to a perfect fluid with positive pressure and equation-of-state parameter $0 < w \leq 1$, leading to
\begin{equation}
c_s^2 < 0 \,\,\, {\rm if} \,\,\,  b^2 < t^2 < \frac{3(2-\alpha)}{2-3\alpha}b^2; \quad c_s^2 < 1 \,\,\,  {\rm if} \,\,\, t^2 > \frac{3(2-\alpha)}{2-3\alpha}b^2, \,\,\, \text{and} \,\,\, c_s^2 > 1\ {\rm if}\,\, t^2 < b^2.
\label{sound6} \end{equation}
This particular case, namely $\frac13 \leq \alpha < \frac{2}{3}$, is presented graphically below in Figure~\ref{fig:sound}.

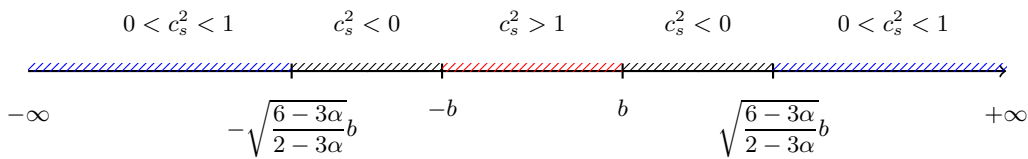
\begin{figure}[th!] \centering
\begin{tikzpicture}[mydrawstyle/.style={draw=black, thick}, x=1mm, y=1mm, z=1mm]
\draw[mydrawstyle, ->](-80,30)--(50,30) node[below=10]{$ +\infty$};
\draw[mydrawstyle](-80,30)--(-80,30) node[below=10]{$-\infty$};
\draw[mydrawstyle](-60,30)--(-60,30) node[above=10]{$0<c_s^2<1$};
\draw[mydrawstyle](-45,29)--(-45,31) node[below=10]{$-\sqrt{\dfrac{6-3\alpha}{2-3\alpha}}b$};
\draw[mydrawstyle](-35,30)--(-35,30) node[above=10]{$c_s^2<0$};
\draw[mydrawstyle](-25,29)--(-25,31) node[below=10]{$-b$};
\draw[mydrawstyle](-13,30)--(-13,30) node[above=10]{$c_s^2>1$};
\draw[mydrawstyle](-1,29)--(-1,31) node[below=10]{$b$};
\draw[mydrawstyle](9,30)--(9,30) node[above=10]{$c_s^2<0$};
\draw[mydrawstyle](35,30)--(35,30) node[above=10]{$0<c_s^2<1$};
\draw[mydrawstyle](19,29)--(19,31) node[below=10]{$\sqrt{\dfrac{6-3\alpha}{2-3\alpha}}b$};
\fill[pattern=north east lines,pattern color=blue] (-80,30) rectangle (-45,31);
\fill[pattern=north east lines] (-45,30) rectangle (-25,31);
\fill[pattern=north east lines, pattern color=red] (-25,30) rectangle (-1,31);
\fill[pattern=north east lines] (-1,30) rectangle (19,31);
\fill[pattern=north east lines, pattern color=blue] (19,30) rectangle (50,31);  
\end{tikzpicture}
\caption{The corresponding squared speed of sound to the possible transformations in the tachyon model~\eqref{tach}, which is shown in Figure~\ref{fig:transitions}.}
\label{fig:sound} \end{figure}
%\end{center}
Finally, in the case of $0 < \alpha < \frac13$, suited for the equation of state with $w > 1$, we have 
\begin{equation}\begin{gathered}
c_s^2 < 0 \quad {\rm if} \quad b^2 < t^2 < \frac{3(2-\alpha)}{2-3\alpha}b^2, \qquad  \text{and} 
\label{sound11} \\ %\end{equation} and \begin{equation}
c_s^2 > 1 \,\,\,  {\rm if} \,\,\, t^2 < b^2 \,\,\, \text{or} \,\,\, \frac{3(1-\alpha)}{1-3\alpha}b^2 < t^2; \quad 
%\label{sound9} \end{equation} \begin{equation}
c_s^2 < 1 \,\,\, {\rm if} \quad \frac{3(2-\alpha)}{2-3\alpha}b^2 < t^2 < \frac{3(1-\alpha)}{1-3\alpha}b^2.
%\label{sound10}  
\end{gathered} \end{equation}%\end{equation} %which can be drawn schematically as}

{Let us now switch off the regularization, i.e., set $b=0$. As follows from Equation~\eqref{sound1},
\begin{equation}
c_s^2 = \frac{2-3\alpha}{3\alpha};
\label{sound12}
\end{equation}
the speed of sound squared is positive if $\alpha < \frac23$, i.e., if the pressure is positive, and it is subluminal if $\alpha > \frac13$, i.e., the pressure is smaller than the energy density. We have seen that in any case, the inclusion of the parameter $b$ and regularizing the metric introduces instabilities into the cosmological solutions. Such a situation looks rather natural. One can remember that, for example, in a very simple cosmological model of a closed Friedmann universe filled with a minimally coupled scalar field, for which the potential includes only a massive term that is quadratic in field, %Please ensure meaning has been retained. 
there are solutions with bounces, but they are actually unstable. This model was studied in detail by many authors; see, e.g., refs. \cite{closed,closed1,closed2,closed3,closed4,closed5,closed6,closed7}. Thus, it looks like a very challenging task to obtain a cosmological model with non-singular, stable~evolution.}

{One can ask themselves: what can be the value of the regularizing parameter $b$? As a matter of fact, because of the purely theoretical nature of our toy model, it is difficult to make some reasonable estimations. One can say only that any, even the most tiny nonzero value, of $b$ does the job of %Please ensure meaning has been retained. 
eliminating the cosmological singularity. On the other hand, the smaller the value of $b$, the less distorting its effects are on other aspects of cosmological evolution. Thus, with a more complicated and realistic nonsingular cosmological model, one can hope to find a bound from above on the values of regularizing parameters when comparing the model with observational data.}

%%%%%%%%%%%%%%%%%%%%%%%%%%%%%%%%%%%%%%%%%%
\section{Discussion}

We applied a simple procedure for the construction of cosmological models free from singularities to flat Friedmann universes filled with minimally coupled scalar fields or by tachyon Born--Infeld-type fields. The form of the regular metric for the Friedmann universes, which we have used in the paper \cite{we-reg-cosm} and in the present paper, was inspired by the prescription used in the paper \cite{Simpson-Visser} for the construction of regular black holes.
Remarkably, for both cases{---}the minimally coupled scalar field and the tachyon field{---}the regularity of the cosmological evolution, or in other words, the existence of bounce, implies the necessity of the transition between scalar fields with standard kinetic terms to those with %Please ensure meaning has been retained. 
phantom ones. In both cases, the potentials of the minimally coupled scalar field and the tachyon in the vicinity of the point of the transition have a non-analyticity of the cusp form that is characterized by the same exponent and is %Please ensure meaning has been retained. 
equal to $\frac23$. If in the tachyon model we choose the evolution such that the pressure changes its sign, then another transformation of the Born--Infeld-type field occurs: the tachyon transforms into a pseudotachyon, and vice versa. 

It is worth noting that a transition between these two types of scalar fields was also investigated in the articles \cite{Vikman:2004,we4,we5} in a rather different context. The starting point there was the observation that the equation of the state of effective dark energy models in the late universe can change its form across the value $w=-1$. This phenomenon is called the ``crossing of the phantom divide line'' in the literature. Onward, the authors of \cite{we4}, inspired by \cite{Yurov}, proposed a model wherein this effect is realized in the presence of a single scalar field; see also the earlier work \cite{Vikman:2004}. For this to be achieved in \cite{we4}, it was necessary to have a cusp in the potential of the scalar field, and its initial conditions needed to be chosen in a special way. Further details of this model were explored in \cite{we5}. Remarkably, the form of the cusp found in \cite{we4} coincides with that found in \cite{we-reg-cosm} for a minimally couple scalar field and, in the present paper, for a tachyon field. Enigmatically for us, the exponent $\frac23$ arises everywhere.

{We would like to mention some other curious facts concerning Born--Infeld-like fields. First, as was noticed in the paper \cite{Frol-Kof-Star}, a cosmological model with a tachyon with constant potential exactly coincides with that based on the Chaplygin gas~\cite{Chap} with an equation of state $p = -A/\rho $, 
%\begin{equation} p = -\frac{A}{\rho}, \label{Chap} \end{equation}
where $A$ is a positive constant; see also ~\cite{Chap1,Chap2,Chap3}. An analogous observation was made in~\cite{we-tach}: a model based on a pseudotachyon with constant potential is equivalent to a model based on a perfect fluid, which was called ``anti-Chaplygin gas'' and has an %Please ensure meaning has been retained. 
equation of state $p = B /\rho$, 
%\begin{equation} p = \frac{B}{\rho}, \label{Chap1} \end{equation}
where $B$ is a positive constant. Remarkably, an equation of state of this type was obtained from the so-called wiggly strings~\cite{Vil,Carter}. The anti-Chaplygin gas appears to be a rather convenient tool for studying future soft singularities.}

Concluding the paper, we would like to say that the study of regular cosmological models free of singularities, just like the investigation of regular black holes, brings some interesting results and reveals some unusual features of General Relativity and its modifications and generalizations. However, eliminating the singularities rather often implies the appearance of some cumbersome and not quite natural types of matter. Thus, in our opinion, the idea that the singularities in General Relativity are not its drawback but its distinguishing feature, which should be accepted and for which adequate language for their treatment should be developed, is very attractive. We complete our text with a reference to an old paper by Charles Misner in which this idea was expressed in a very clear and convincing way \cite{Misner1}.  
In particular, he wrote, ``We should stretch our minds, find some more acceptable set of words to describe the mathematical situation, now identified as `singular', and then proceed to incorporate this singularity into our physical thinking until observational difficulties force revision on us.
The concept of a true initial singularity (as distinct from an indescribable early era at extravagant but finite high densities and temperatures) can be a positive and useful element in cosmological theory.''

\acknowledgments{This research was partially supported by the INFN grant FLAG.}

\end{document}